\documentclass[usenatbib]{mn2e}

\usepackage{amsmath, amssymb}
\usepackage{array}
\usepackage{graphicx}

\begin{document}

\newcommand{\be}{\begin{equation}}
\newcommand{\ee}{\end{equation}}
\newcommand{\pdf}{p}
\newcommand{\data}{{d}}
\newcommand{\Df}{{D}}
\newcommand{\Bf}{\mathcal{B}_{01}}
\newcommand{\lnBf}{\ln\Bf}
\newcommand{\mdl}{{M}}
\newcommand{\lsim}{\,\raise 0.4ex\hbox{$<$}\kern -0.8em\lower 0.62ex\hbox{$\sim$}\,}
\newcommand{\gsim}{\,\raise 0.4ex\hbox{$>$}\kern -0.7em\lower 0.62ex\hbox{$\sim$}\,}

\newcommand{\params}{{\theta}}
\newcommand{\mean}{{\mu}}
\newcommand{\like}{L}
\newcommand{\lnlike}{\mathcal{L}}
\newcommand{\ML}{^*}
\newcommand{\dr}{\textrm{d}}
\newcommand{\ie}{i.e.}
\newcommand{\reion}{\text{re}}

\newcommand{\cd}{\cdot}
\newcommand{\cds}{\cdots}
\newcommand{\ip}{\int_0^{2\pi}}
\newcommand{\al}{\alpha}
\newcommand{\ba}{\beta}
\newcommand{\de}{\delta}
\newcommand{\De}{\Delta}
\newcommand{\ep}{\epsilon}
\newcommand{\Ga}{\Gamma}
\newcommand{\ka}{\tau}
\newcommand{\io}{\iota}
\newcommand{\La}{\Lambda}
\newcommand{\Om}{\Omega}
\newcommand{\om}{\omega}
\newcommand{\si}{\sigma}
\newcommand{\Si}{\Sigma}
\newcommand{\te}{\theta}
\newcommand{\ze}{\zeta}
\newcommand{\vth}{\ensuremath{\vartheta}}
\newcommand{\vph}{\ensuremath{\varphi}}
\newcommand{\MM}{\mbox{$\cal M$}}
\newcommand{\tr}{\mbox{tr}}
\newcommand{\hor}{\mbox{hor}}
\newcommand{\grad}{\mbox{grad}}
\newcommand{\cx}{\ensuremath{\mathbf{\nabla}}}
\newcommand{\lap}{\triangle}
\newcommand{\arctg}{\mbox{arctg}}
\newcommand{\bm}[1]{\mbox{\boldmath $#1$}}
\newcommand{\eff}{{\rm eff}}
\newcommand{\tto}{\Rightarrow}
\newcommand{\lag}{\langle}
\newcommand{\rag}{\rangle}
\newcommand{\fiso}{f_{\text{iso}}}
\newcommand{\Afiso}{\vert f_{\text{iso}}\vert}
\newcommand{\norm}[3]{{N_{#1,#2}(#3)}}
\newcommand{\fexp}{\boldsymbol{e}}
\newcommand{\KL}{D_{KL}}
\newcommand{\ns}{n_S}
\newcommand{\omk}{\Om_\kappa}
\newcommand{\eps}{\sigmailon}
\title[Forecasting the Bayes factor]{Forecasting the Bayes factor of a future observation}

\author[Roberto Trotta]{Roberto Trotta\thanks{E-mail address: {\tt rxt@astro.ox.ac.uk}}
\\
Oxford University, Astrophysics,  Denys Wilkinson Building, Keble
Road, OX1 3RH, United Kingdom}

\maketitle

\begin{abstract}
I present a new procedure to forecast the Bayes factor of a future
observation by computing the Predictive Posterior Odds
Distribution (PPOD). This can assess the power of future
experiments to answer model selection questions and the
probability of the outcome, and can be helpful in the context of
experiment design.

As an illustration, I consider a central quantity for our
understanding of the cosmological concordance model, namely the
scalar spectral index of primordial perturbations, $n_S$. I show
that the Planck satellite has over $90\%$ probability of gathering
strong evidence against $n_S = 1$, thus conclusively disproving a
scale--invariant spectrum. This result is robust with respect to a
wide range of choices for the prior on $n_S$.
\end{abstract}



\section{Introduction}

Many interesting questions in cosmology are not about parameter
estimation, but rather about model selection. For example, we
might be interested in assessing whether a new parameter is needed
in our model, or whether a theoretical prediction for the value of
a parameter can be confirmed by data.

These kind of questions often cannot be satisfactorily answered in
the context of frequentist (sampling theory) statistics, but find
their natural formulation in the framework of Bayesian model
selection (see \cite{Trotta:2005ar,Liddle:2007fy} and references
therein). Bayesian model selection aims at working out the support
that the data can offer to a model, by balancing the quality of
fit that a more complicated model usually delivers with a
quantitative embodiment of Occam's razor, favouring simpler
explanations whenever they are compatible with the observations at
hand. This is usually expressed in terms of the Bayes factor
between two competing models, which represents the amount by which
our relative believe in the two model has changed after the
arrival of the data. There is a growing body of work in cosmology
and astrophysics applying various brands of model selection tools
to a broad range of questions, see
e.g.~\cite{Drell:1999dx,Loredo:2001rx,Hobson:2002de,Slosar:2002dc,Saini:2003wq,Lazarides:2004we,Beltran:2005xd,Kunz:2006mc,Marshall:2004zd,
Magueijo:2006we,Parkinson:2006ku,Trotta:2006ww,Trotta:2006pw,Bevis:2007gh}.

The purpose of this paper is to present a new method to forecast
the probability distribution of the Bayes factor for a future
observation, called PPOD (for ``Predictive Posterior Odds
Distribution'')\footnote{The method was called ExPO (for
``Expected Posterior Odds'') in a previous version of this
work~\citep{Trotta:2005ar_v1}. I am grateful to Tom Loredo for
suggesting the new, more appropriate name.}.

Posterior odds forecasting was first introduced in
\cite{Trotta:2005ar_v1}, which used a single model to describe the
present data. This has inspired further developments of a similar
technique in \cite{Pahud:2006kv,Pahud:2007gi}. In particular,
\cite{Pahud:2006kv} pointed out that the Bayes factor forecasting
ought to consider multiple models and average over them. This
approach is used in the present work. For a different approach to
Bayes factor forecasting, see \cite{Mukherjee:2005tr}, which
instead focuses on delineating regions of parameter space where
future observations have the ability of delivering high--odds
model selection results.

The use of our PPOD technique is illustrated on a central
parameter of the cosmological concordance model, namely the scalar
spectral index for cosmological perturbations, $n_S$, which can be
related to the characteristics of the inflationary potential (see
e.g.~\cite{Leach:2003us}). One interesting question bears on
whether the distribution of fluctuations is scale--invariant,
i.e.~whether a model with $n_S = 1$ (the so-called
Harrison-Zeldovich power spectrum) is supported by data. Current
cosmological observations support the view that $n_S \neq 1$, with
odds of about $17:1$ \citep{Trotta:2005ar} (see also
\cite{Pahud:2006kv}, who find odds of $8:1$ in favour of $n_S \neq
1$). In this paper, we derive a predictive distribution for $n_S$
for the Planck satellite -- an European cosmic microwave
background satellite due for launch next year -- and present a
forecast for the model selection outcome from Planck observations.

This paper is organized as follows: in section \ref{sec:Bayesian}
we briefly review the main concepts of Bayesian model comparison.
We then introduce our PPOD technique in section \ref{sec:expod}
and we apply it to derive the probability distribution for the
model selection outcome from Planck in section
\ref{sec:application}, also discussing the dependence on the
choice of prior. Section \ref{sec:conclusions} is devoted to
presenting our conclusions.

\section{Bayesian model comparison}

\label{sec:Bayesian}

In this section we briefly review Bayesian model comparison and
introduce our notation.

Bayesian inference (see e.g.~\cite{JaynesBook,MKbook}) is based on
Bayes' theorem, which is a consequence of the product rule of
probability theory:
 \be \label{eq:Bayes_Theorem}
    p(\params | \data,\mdl)= \frac{p(\data | \params,\mdl)
    \pi(\params \vert \mdl)}{p(\data | \mdl)}.
\ee On the left-hand side, the posterior probability for the
parameters $\params$ given the data $\data$ under a model $\mdl$
is proportional to the likelihood $p(\data | \params,\mdl)$ times
the prior probability distribution function (pdf), $\pi(\params |
\mdl)$, which encodes our state of knowledge before seeing the
data. In the context of model comparison it is more useful to
think of $\pi(\params | \mdl$) as an integral part of the model
specification, defining the prior available parameter space under
the model $\mdl$~\citep{Kunz:2006mc}. The normalization constant
in the denominator of \eqref{eq:Bayes_Theorem} is the {\em
marginal likelihood for the model $\mdl$} (sometimes also called
the ``evidence'') given by
 \be \label{eq:evidence_def}
 p(\data | \mdl) = {\int_{\Omega} p(\data|\params,M)
\pi(\params | \mdl)\dr\params}
 \ee
where $\Omega$ designates the parameter space under model $\mdl$.
In general, $\params$ denotes a multi--dimensional vector of
parameters and $\data$ a collection of measurements.

Consider two competing models $\mdl_0$ and $\mdl_1$ and ask what
is the posterior probability of each model given the data $\data$.
By Bayes' theorem we have
\begin{equation}
\pdf(\mdl_i\vert\data)\propto
\pdf(\data\vert\mdl_i)\pi(\mdl_i)~~(i=0,1),
\end{equation}
where $\pdf(\data\vert\mdl_i)$ is the marginal likelihood for
$\mdl_i$ and $\pi(\mdl_i)$ is the prior probability of the $i$th
model before we see the data. The ratio of the likelihoods for the
two competing models is called the {\it Bayes factor}:
\begin{equation} \label{eq:bfac}
B_{01}\equiv\frac{\pdf(\data\vert\mdl_0)}
{\pdf(\data\vert\mdl_1)},
\end{equation}
which is the same as the ratio of the posterior probabilities of
the two models in the usual case when the prior is presumed to be
noncommittal about the alternatives and therefore
$\pi(\mdl_0)=\pi(\mdl_1)=1/2$. The Bayes factor can be interpreted
as an automatic Occam's razor, which disfavors complex models
involving many parameters (see e.g.~\cite{MKbook} for details, as
well as the discussion in~\cite{rebuttal}). A Bayes factor $B_{01}
> 1$ favors model $\mdl_0$ and in terms of betting odds it would
prefer $\mdl_0$ over $\mdl_1$ with odds of $B_{01}$ against 1. The
reverse is true for $B_{01} < 1$.

It is usual to consider the logarithm of the Bayes factor, for
which the ``Jeffreys' scale'' for the strength of evidence offers
an empirically calibrated rule of thumb~\citep{Jeffreys,Kass}.
Different authors use different conventions to describe the
strength of evidence -- in this work we use the same convention of
\cite{Trotta:2005ar}, deeming values  $\vert \ln B_{01} \vert > 1;
> 2.5; > 5.0$ to constitute `positive', `moderate' and `strong'
evidence, respectively.

Evaluating the marginal likelihood integral of
Eq.~\eqref{eq:evidence_def} is in general a computationally
demanding task for multi--dimensional parameter spaces. Several
techniques are available on the market, each with its own
strengths and weaknesses: thermodynamic
integration~\citep{Slosar:2002dc,Beltran:2005xd}, nested sampling
(introduced by \cite{SkillingNS} and implemented in the
cosmological context by~\cite{Bassett:2004wz,Mukherjee:2005wg}),
or the Savage--Dickey density ratio (SDDR), introduced
in~\cite{Trotta:2005ar}. Since the method present here makes use
of the SDDR, we briefly remind the reader about it, referring
to~\cite{Trotta:2005ar} for further details.

If we wish to compare a two--parameters model $\mdl_1$ with a
restricted submodel $\mdl_0$ with only one free parameter, $\psi$,
and with fixed $\omega = \omega_\star$ and assuming further that
the prior is separable (which is usually the case in cosmology),
i.e.\ that
 \begin{equation}
 \pi(\omega, \psi | \mdl_1) = \pi(\omega | \mdl_1) \pi(\psi |
 \mdl_0),
 \end{equation}
then the Bayes factor $B_{01}$ of Eq.~\eqref{eq:bfac} can be
written as
 \begin{equation} \label{eq:savagedickey}
 B_{01} = \left.\frac{\pdf(\omega \vert \data, \mdl_1)}{\pi(\omega |
 \mdl_1)}\right|_{\om = \om_\star} \quad {\text{(SDDR)}}.
 \end{equation}
Thanks to the SDDR, the evaluation of the Bayes factor of two
nested models only requires the properly normalized value of the
marginal posterior at $\omega = \omega_\star$ under the extended
model $\mdl_1$, which is a by--product of parameter inference. We
note that the derivation of \eqref{eq:savagedickey} {\it does not
involve any assumption about the posterior distribution}, and in
particular about its normality. As it has been shown in Appendix C
of~\cite{Trotta:2005ar}, the SDDR works well if the parameter
value under the simpler model, $\om_\star$, is not too far away
from the mean of the posterior under the extended model. The
reason for this is that it becomes increasingly cumbersome to
reconstruct the posterior with enough accuracy  in the tails of
the distribution. More specifically, for distributions close to
Gaussian, Eq.~\eqref{eq:savagedickey} is likely to be reliable if
$\om_\star$ is less than about 3 standard deviations away from the
mean of the posterior.

We now turn to describing our forecast technique allowing to
obtain a probability distribution for the Bayes factor from future
observations.

\section{Bayes factor forecast: PPOD} \label{sec:expod}

In designing a new observation, it is interesting to assess its
potential in terms of its power to address model comparison
questions. To this end, we introduce a new technique which
combines a Fisher information matrix forecast with the SDDR
formula to obtain a forecast for the Bayes factor of a future
observation. The result is a PPOD (for {\em ``Predictive Posterior
Odds Distribution''}) for the future model comparison results.

\subsection{The predictive distribution}

We are interested in predicting the distribution of future data,
from which the result of a future model comparison can be
obtained. The {\em predicting distribution} for future data $\Df$
is
 \be \label{eq_pred_data}
 \begin{aligned}
 \pdf(\Df|\data) & = \sum_{i=0}^1
 \pdf(\Df|\data,\mdl_i)\pdf(\mdl_i|d) \\
 & = \sum_{i=0}^1 \pdf(\mdl_i|d) \int
 p(\Df|\params,\mdl_i)p(\params|\data,\mdl_i) \dr\params,
 \end{aligned}
 \ee
 where the sum runs over the 2 competing models we are
considering\footnote{An earlier version of this work did not carry
out the sum over models, but was restricted to the $i=1$ term of
Eq.~\eqref{eq_pred_data}. I am grateful to Andrew Liddle for
bringing this to my attention. This is also spelled out in
\cite{Pahud:2006kv}.}. Generalization to a larger number of models
is straightforward. In the above, $p(\Df|\params,\mdl_i)$ is the
predicted likelihood for future data, assuming $\params$ is the
correct value for the cosmological parameters (under model
$\mdl_i$). A Gaussian approximation to the future likelihood can
be obtained by performing a Fisher Matrix analysis (FMA) assuming
$\params$ as a fiducial model. This yields a forecast of the
parameters covariance matrix ${\bf C}$ for future data $\Df$ (for
a detailed account, see
e.g.~\cite{Knox:1995dq,Kosowsky:1996mp,Efstathiou:1998xx,Rocha:2004}).

The corresponding predictive posterior odds distribution (PPOD)
for the future Bayes factor, $\Bf$, conditional on current data
$\data$ is then
 \be \label{eq;future_PPOD}
 \begin{aligned}
 \pdf(\Bf|\data) & = \int \pdf(\Bf, \Df|\data)\dr\Df =
 \int\pdf(\Bf|\Df,\data)\pdf(\Df|\data)\dr\Df \\
  &= \int \delta(\Df - \Bf(\Df))\pdf(\Df|\data)\dr\Df
  \end{aligned},
 \ee
 where $\delta$ denotes the Dirac delta--function, and $\Bf(\Df)$
denotes the functional relationship between future data and the
Bayes factor, given in our case by the SDDR,
Eq.~\eqref{eq:savagedickey}. The presence of the delta--function
comes from the univocal relationship between the future data and
$\Bf$ (see Eq.~\eqref{eq:max_B01} below). In other words, the
Bayes factor is simply a derived parameter of the future
likelihood.

It is instructive to consider the Gaussian case, whose PPOD can be
written down analytically. We restrict ourselves to the case of
nested models, and we write for the parameter space of the
extended model $\params = \left(\omega , \psi \right)$, where
$\om$ denotes the extra parameter. If the predicted likelihood
covariance matrix $\bf C$ does not depend on $\psi$ (in other
words, if the future errors do not depend on the location in the
subspace of parameters common to both models), it is easy to see
from Eq.~\eqref{eq_pred_data} that one can marginalize over the
parameters common to both models, $\psi$. Thus we can assume
without loss of generality a 1--dimensional $\mdl_1$ compared with
model $\mdl_0$ with no free parameters. We take a Gaussian prior
on the extra parameter, centered around 0 and of width equal to
unity (this can always been achieved by suitably rescaling and
shifting the variables), that we denote by
 \be \label{eqppod:prior}
  \pdf(\om|\mdl_1) = \norm{0}{1}{\om}
 \ee
and describe the present--day likelihood as a Gaussian centered on
$\om=\mu$ of width $\sigma$, where $(\mu,\sigma)$ are understood
to be expressed in units of the prior width and are thus
dimensionless:
  \be \label{eqppod:present_like}
  \pdf(\data|\om, \mdl_1) = \norm{\mu}{\sigma}{\om}
 \ee
The predicted likelihood under future data $\Df$ is also Gaussian
distributed, with mean $\om=\nu$ and (constant) standard deviation
$\tau$:
  \be \label{eqppod:future_like}
  \pdf(\Df|\om, \mdl_1) = \norm{\nu}{\tau}{\om}.
 \ee
Here, the forecasted error $\tau = \sqrt{{\bf C}_{11}}$ is taken
to be independent on $\om$, and is understood to be the marginal
error on $\om$, after marginalizing over the common parameters
$\psi$. Using Eqs.~(\ref{eqppod:prior}--\ref{eqppod:future_like})
into \eqref{eq_pred_data} we obtain after a straightforward
calculation
 \be \label{eq:ppod_future_pred}
 \begin{aligned}
 \pdf(\Df|\data) \propto
 & \frac{\pdf(\mdl_0)}{\tau\sigma}\exp\left(-\frac{1}{2}\frac{\nu^2\sigma^2+\mu^2\tau^2}{\tau^2\sigma^2}
 \right)+ \\
 & \frac{\pdf(\mdl_1)}{\sqrt{\tau^2+\sigma^2+\tau^2\sigma^2}}
 \exp\left(-\frac{1}{2}\frac{(\nu-\mu)^2 + \sigma^2\nu^2 + \tau^2\mu^2}{\tau^2+\sigma^2+\tau^2\sigma^2}
 \right),
 \end{aligned}
 \ee
where we have dropped irrelevant constants. As a function of the
future mean $\nu$, Eq.~\eqref{eq:ppod_future_pred} gives the
probability of obtaining a value $\om=\nu$ from a future
measurement, conditional on the present data $\data$ and on the
current model selection outcome. The PPOD can be obtained from
\eqref{eq;future_PPOD} and \eqref{eq:ppod_future_pred} by using
the relation between $\nu^2$ and $\ln\Bf$ (obtained by applying
the SDDR):
 \be \label{eq:max_B01}
 \nu^2 =
 \tau^2(1+\tau^2)\left(\ln\frac{1+\tau^2}{2\pi\tau^2}
 - 2 \ln\Bf\right)
 \ee
For $\nu = 0$, corresponding to the future observation measuring
the predicted value of $\om$ under $\mdl_0$,
Eq.~\eqref{eq:max_B01} gives the maximum odds in favour of model
$\mdl_0$ one can hope to gather from a future measurement with
error $\tau$.

In the general case, where the current likelihood is non--Gaussian
and the future likelihood covariance matrix can depend on
$\params$, it is possible to compute $p(\Df|\data)$ numerically
from a series of MCMC samples. By using a similar manipulation as
the one illustrated in Appendix B of~\cite{Trotta:2005ar} to
obtain the SDDR formula, we can recast the $i=0$ term in the
sum~\eqref{eq_pred_data} as
 \be \label{eq:manip_P0}
 \begin{aligned}
& \pdf(\mdl_0|d) \int
  \pdf(\Df|\psi,\mdl_0)\pdf(\psi|\data,\mdl_0) \dr\psi = \\
&  B_{01}\frac{\pdf(\mdl_0)}{\pdf(\data)} \int
 \pdf(\Df|\psi,\mdl_0)\pdf(\data|\psi,\om_\star, \mdl_1)\pdf(\psi,\om_\star|\mdl_1) \dr\psi.
\end{aligned}
 \ee
Since the constant factor $\pdf(\data)^{-1}$ is common to both
terms in the sum and hence factors out, knowledge of the
un--normalized posterior under $\mdl_1$ and of the present--day
Bayes factor $B_{01}$ is sufficient to compute the predictive data
distribution and therefore the PPOD by employing
Eq.~\eqref{eq:max_B01}. Given $N$ independent samples from the
un--normalized posterior under $\mdl_1$, $\pdf(\data|\psi,\om,
\mdl_1)\pdf(\psi,\om|\mdl_1)$, which can be obtained by standard
MCMC techniques, one proceeds to perform a FMA at every sample,
thus obtaining a prediction for the future covariance matrix at
that point in parameter space. Let us denote the MCMC samples by
$\params_j = (\om_j, \psi_j)$, $j=1,\dots,N$. The predictive data
distribution \eqref{eq_pred_data} is obtained by averaging the
future likelihood over the samples, i.e. using
Eq.~\eqref{eq:manip_P0}
 \be \label{eq:pred_distr_chains}
 \begin{aligned}
 \pdf(\Df|\data) \propto & \pdf(\mdl_1)\frac{1}{N}\sum_{j=1}^N
 \pdf(\Df|\om_j,\psi_j,\mdl_1) + \\
 & \pdf(\mdl_0)B_{01}\frac{1}{K}\sum_{k=1}^K
 \pdf(\Df|\om_\star,\psi_k,\mdl_1)
\end{aligned}
 \ee
where $K$ is the number of samples in the chain with
$\om=\om_\star$ (or within a suitably small neighbourhood from
$\om_\star$) and we have dropped an overall normalization factor
$\pdf(\data)^{-1}$. The corresponding PPOD for $\Bf$ can then be
obtained using Eqs.~\eqref{eq:max_B01} and \eqref{eq;future_PPOD}.

The predictive distribution of Eq.~\eqref{eq:pred_distr_chains}
does not make any assumptions regarding the normality of the
current posterior, nor of the prior. However, it does assume that
the future likelihood can be described by a Gaussian distribution,
as is implicit in the use of the FMA. This aspect is not so
critical, since FMA errors have proved to give reliable estimates,
especially when using ``normal parameters''
\citep{Kosowsky:2002zt}. The second assumption is hidden in
Eq.~\eqref{eq:max_B01}, which relates the future Bayes factor
$\Bf$ to the future mean, $\nu$. This relation only holds for a
Gaussian prior and assuming that the posterior for future data is
accurately described by a Gaussian, which is likely to break down
in the tails of the distribution, $|\nu - \om_\star|/\tau \gg 1$.
Nevertheless, we can still conclude that models which have $|\nu -
\om_\star|/\tau \gg 1$ strongly disfavor $\mdl_0$ under future
data, even though we cannot attach a precise value to the expected
odds. This is why we present PPOD results by giving only the
integrated probability within a few coarse regions, as in
Table~\ref{tab:PPOD}. We notice that one could improve on both of
the above assumptions by using MCMC techniques to sample from the
future likelihood rather than using a Gaussian approximation. This
however would add considerably to the computational burden of the
forecast.

\subsection{Extension to experiment design}

Our approach can be extended to the context of Bayesian experiment
design, whose goal is to optimize a future observation in order to
achieve the maximum science return (often defined in terms of
information gain or through a suitable figure of merit, see
\cite{Loredo:2003nm} and references therein for an overview,
\cite{Bassett:2004st,Bassett:2004np,Parkinson:2007cv} for a more
cosmology--oriented application and \cite{Ford:2004gu} for an
astrophysical application).

The core of the procedure is the quantification of the utility of
an experiment as a function of the experimental design, possibly
subject to experimental constraints (such as observing time,
sensitivity, noise characteristics, etc). The observing strategy
and experiment design are then optimized to maximise the expected
utility of the observation. The PPOD is a good candidate for an
utility function aimed at model comparison, for it indicates the
probability of reaching a clear--cut model distinction thanks to
the future observation. The dependence on experimental design
parameters is implicit in the FMA, and therefore one could imagine
optimizing the choice of experimental parameters to maximise the
probability of obtaining large posterior odds from the future
data, integrating over current posterior knowledge. This is
especially interesting since it marginalizes over our current
uncertainty in the value of the parameters, rather than assuming a
fiducial model as it is usually done in Fisher matrix forecasts
common in the literature.

Since in the present paper we focus on model comparison rather
than experiment design, in the following we fix the experimental
parameters for the Planck satellite to the value used in
\cite{Rocha:2004}. We leave further exploration of the issue of
design optimization and PPOD for future work.

\section{Forecasts for the Planck satellite}
 \label{sec:application}

In this section we investigate the potential of the Planck
satellite in terms of model comparison results. For other works
using a similar technique, partially inspired by our approach,
see~\cite{Pahud:2006kv,Pahud:2007gi}

\subsection{Parameter space and current cosmological data}

As current cosmological data, we use the WMAP 3--year temperature
and polarization data \citep{Hinshaw:2006ia,Page:2006hz}
supplemented by small--scale CMB measurements
\citep{Readhead:2004gy,Kuo:2002ua}. We add the Hubble Space
Telescope measurement of the Hubble constant $H_0 = 72 \pm 8$
km/s/Mpc \citep{Freedman:2000cf} and the Sloan Digital Sky Survey
(SDSS) data on the matter power spectrum on linear scales ($k <
0.1 h^{-1}\rm Mpc$)~\citep{Tegmark:2003uf}. Furthermore, we shall
also consider supernovae luminosity distance
measurements~\citep{Riess:2004nr}. We make use of the publicly
available codes CAMB and CosmoMC \cite{Lewis:2002ah} to compute
the CMB and matter power spectra and to construct Monte Carlo
Markov Chains (MCMC) in parameter space. We sample uniformly over
the physical baryon and cold dark matter (CDM) densities,
$\omega_b \equiv \Omega_b h^2$ and $\omega_c\equiv\Omega_c h^2$,
expressed in units of $1.88\times 10^{-29}~{\rm g/cm}^3$; the
ratio of the angular diameter distance to the sound horizon at
decoupling, $\Theta_\star$, the optical depth to reionization
$\tau_r$ (assuming sudden reionization) and the logarithm of the
adiabatic amplitude for the primordial fluctuations, $\ln 10^{10}
A_S$. When combining the matter power spectrum with CMB data, we
marginalize analytically over a bias $b$ considered as an
additional nuisance parameter. Throughout we assume three massless
neutrino families and no massive neutrinos, we neglect the
contribution of gravitational waves to the CMB power spectrum and
we assume a flat Universe.

\subsection{PPOD forecast for the spectral index}

From the current posterior we can produce a PPOD forecast for the
Planck satellite\footnote{See the website:
\texttt{http://astro.estec.esa.nl/Planck} .}  following the
procedure outlined in Section \ref{sec:expod}. As motivated in the
introduction, we focus on the scalar spectral index $n_S$ and we
follow the same setup as in \cite{Trotta:2005ar}, comparing an
Harrison--Zeldovich (HZ) $n_S = 1$ model against a generic
inflationary model with a Gaussian prior of width $\Delta n_S =
0.2$, as motivated by slow--roll inflation. In
\cite{Trotta:2005ar} it was shown that a compilation of
present--day CMB, large scale structure, supernovae and Hubble
parameter measurements yields moderate odds ($17:1$) in favour of
$n_S\neq 1$.

\begin{figure}
\centering
\includegraphics[width=\linewidth]{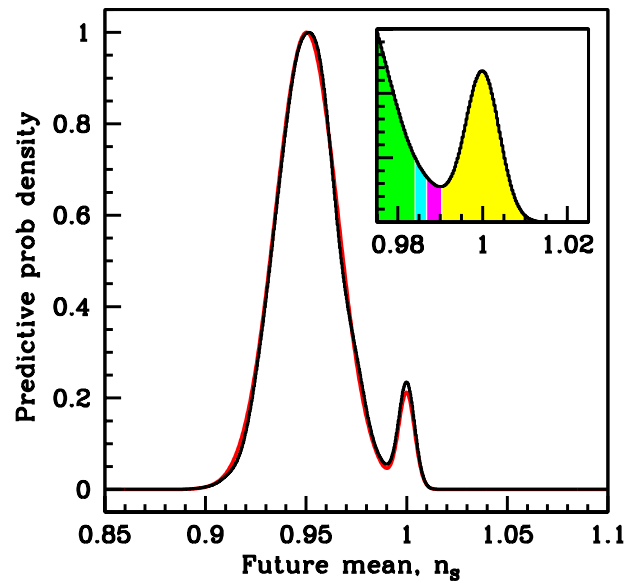}
\caption{Predictive data distribution for the Planck satellite
nominal mission, conditional on current (WMAP3+ext) knowledge. We
are plotting the probability distribution (normalized to the peak)
of the future measurement of the spectral tilt. The bump at $n_S =
1$ corresponds to the probability associated with the HZ model.
The black curve is obtained numerically from MCMC chains
(Eq.~\eqref{eq:pred_distr_chains}) while the red is for the
Gaussian approximation, Eq.~\eqref{eq:ppod_future_pred}. In the
inset, the shaded areas delimit regions where the Bayes factor
from Planck deliver strong evidence in favour of $n_S\neq1$
(green, $\lnBf < -5.0$, this region extending to all smaller $n_S$
values), moderate evidence for $n_S\neq1$ (cyan, $-5.0 \leq \lnBf
\leq -2.5$), positive evidence for $n_S\neq1$ (magenta, $-2.5 \leq
\lnBf \leq 0.0$) or favour $n_S = 1$ (yellow, $\lnBf > 0$). The
corresponding probability values are reported in
Table~\ref{tab:PPOD}. } \label{fig:ppodns}
\end{figure}

The result in terms of the predictive data distribution is shown
in Figure~\ref{fig:ppodns} and the corresponding PPOD for the
Bayes factor is given in Table~\ref{tab:PPOD} for our choice of
the prior scale, $\Delta n_S = 0.2$ (see below for a discussion of
the dependence of our results on the prior choice). In
Figure~\ref{fig:ppodns} we plot $\pdf(\Df|\data)$ for Planck
conditional on present--day information both as obtained
numerically from the MCMC chains, via
Eq.~\eqref{eq:pred_distr_chains}, and by using the Gaussian
approximation with constant future errors,
Eq.~\eqref{eq:ppod_future_pred}, with $\sigma= 0.015/\Delta n_S =
0.075$, $\tau=0.004/\Delta n_S=0.02$ and $\mu = -0.05/\Delta n_S =
-0.25$ (all these quantities are expressed in units of the prior
width, $\Delta n_S = 0.2$). We observe that
Eq.~\eqref{eq:ppod_future_pred} is an extremely good approximation
to the full numerical result, obtained from 2000 thinned samples
of a MCMC chain. This follows from the facts that the current
posterior is close to Gaussian, and that the future errors
forecasted for Planck vary only very mildly over the range of
parameter space singled out by the present posterior. Furthermore,
the future errors are almost uncorrelated with the fiducial value
of $n_S$.

We then obtain the PPOD numerically via Eq.~\eqref{eq:max_B01} and
we integrate the distribution to get the probability of the model
comparison result from future data (tabulated in
Table~\ref{tab:PPOD}). The main finding is that Planck has a very
large probability ($\Pr(\lnBf < -5) = 0.928$) to obtain a
high--odds result strongly favouring a spectral tilt over an HZ
spectrum. This is consequence of the fact that the most probable
models under current data are clustered around $n_S = 0.95$ and
that Planck sensitivity will decrease the error around those
models by a factor $\sim 4$. The region of the predictive
distribution corresponding to decisive odds in favour of $n_S \neq
1$ is shown in green in the inset of Figure~\ref{fig:ppodns}, and
it extends to all values $n_S \lsim 0.984$. By contrast, the
probability that Planck will overturn the present model selection
result favouring $n_s \neq 1$ (currently with odds of about
$17:1$, see \cite{Trotta:2005ar}) is only around $6\%$. We also
find that the maximum odds by which Planck could favour $n_S=1$
are of $20:1$, or $\lnBf = 3.00$ (for our choice of prior width),
which would still fall short of the mark of ``strong'' evidence.
It is interesting to note from Table~\ref{tab:PPOD} that either
temperature information or E--polarization information alone will
be enough to deliver a high odds result with large probability
(around $90\%$ in either case).

The above findings are in good agreement with the conclusions in
\cite{Pahud:2006kv}, which were obtained using a more qualitative
version of our procedure. The PPOD procedure presented here
improves on several, potentially important aspects with respect to
the method used in \cite{Pahud:2006kv,Pahud:2007gi}: PPOD takes
into account the full predictive distribution, and in particular
the potentially important tails of the distribution above $n_S=1$;
it fully accounts for the possibility that $n_S = 1$ but that
Planck will actually end up (wrongly) favouring the HZ model
because of a measurement in the tail of the predictive
distribution for $M_0$; finally, it takes into account the effect
due to the variation of the future error on $n_S$ across the
current posterior (even though this aspect has been shown to be
negligible in the present case).

\begin{table}
\caption{Probability of future model comparison results (PPOD) for
the Planck satellite, conditional on present knowledge. There is
about $93\%$ probability that Planck will be able to strongly
favour $n_S\neq1$, i.e. that it will obtain $\lnBf < -5$. The
probability that Planck will favour a scale-invariant spectrum
($\lnBf > 0$) is only about $6\%$. These results are robust even
when using only temperature (TT only column) or only
E--polarization information (EE only column). } \label{tab:PPOD}
\begin{tabular}{l | c c c }
\hline
 \multicolumn{4}{c}{Spectral index: $\ns=1$  versus $0.8
\leq \ns \leq 1.2$ (Gaussian)} \\\hline
             & All & EE only & TT only \\
             $\Pr(\lnBf < -5)$ & $0.928$ & $0.903$ & $0.926$\\
             $\Pr(-5 < \lnBf < -2.5)$ & $0.005$ & $0.018$ & $0.007$ \\
             $\Pr(-2.5 < \lnBf <
             0)$ & $0.006$ & $0.023$ & $0.008$\\
             $\Pr(\lnBf > 0)$ & $0.061$ & $0.056$ & $0.059$\\\hline
\end{tabular}
\end{table}

\subsection{Dependence on the choice of prior}
\label{sec:priors}

\begin{figure}
\centering
\includegraphics[width=\linewidth]{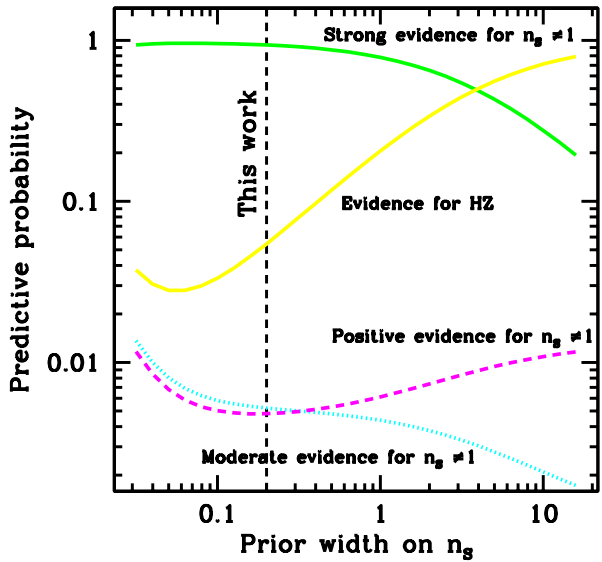}
\caption{PPOD dependence on the prior width for $n_S$. We plot the
PPOD result computed from the Gaussian approximation of
Eq.~\eqref{eq:ppod_future_pred} as a function of the width of the
prior on $n_S$ for the model with $n_S \neq1$. In order to change
the conclusion of this work, namely that Planck has a large
probability of conclusively measuring $n_S \neq 1$, one would have
to adopt a prior larger than about 3.5 (crossing of the green and
yellow lines). In this work, the prior width has been set to 0.2
(dotted, vertical line). } \label{fig:nsprior}
\end{figure}

The prior assignment is an irreducible feature of Bayesian model
selection, as it is clear from its presence in the denominator of
Eq.~\eqref{eq:savagedickey}. In fact, the prior width controls the
strength of the Occam's razor effect on the extended model, and
thus a larger prior favours the simpler model.

We can assess the impact of a change of prior on our PPOD results
by plotting them as a function of the chosen prior width. In
Figure~\ref{fig:nsprior} we show how the probabilities for Planck
to obtain different levels of evidence for or against $n_s \neq 1$
change with a change in the choice of the prior width $\Delta
n_S$. It is apparent that our result holds true for a wide range
of prior values: even if the prior is widened to $\Delta n_S = 1$,
the probability of a strong ($\lnBf < -5$, green line) result in
favour of $n_s\neq1$ is still about $80\%$. The prior width has to
be enlarged to $\Delta n_S \gsim 3.5$ for the simpler model to
have more than 50\% probability of being favoured (yellow line,
depicting the probability of obtaining $\lnBf > 0$).

\section{Conclusions}
\label{sec:conclusions}

We have presented a new statistical technique (PPOD) to produce
forecasts for the probability distribution of the Bayes factor
from future experiments. The use of PPOD can complement the Fisher
matrix forecasts in that it allows to assess the capabilities of a
future experiment to obtain a high--odds model selection result.
Being conditional on present knowledge, our PPOD technique does
not assume a fiducial model, but takes into account the current
uncertainty in the values of the underlying model parameters.

We emphasize that the PPOD forecast, being conditional on the
present posterior, is reliable provided there will be no major
{\em systematic} shift in the parameter determination with respect
to present--day data. In other words, the PPOD only takes into
account the statistical properties of our knowledge, a point
hardly worth highlighting (if we knew the outcome of a future
measurement, it would be pointless to carry it out).

We have applied this method to a central parameter of the
concordance model. We have found that the Planck satellite has
over $90\%$ probability of obtaining a strong ($\lnBf < -5$) model
selection result favouring $n_S\neq1$ (for a prior width $\Delta
n_S = 0.2$), thus improving on current, moderate odds (of about
$17:1$ or $\ln B_{01} = -2.86$). The probability that Planck will
find evidence in favour of $n_S = 1$ is by contrast only about
$6\%$. These results are qualitatively unchanged for a wide range
of prior values, encompassing most reasonable prior choices.

{\em Acknowledgments} I am grateful to Bruce Bassett and Tom
Loredo for useful comments. I thank Martin Kunz, Andrew Liddle and
Pia Mukherjee for helpful suggestions and comments on an early
draft. This research is supported by the Royal Astronomical
Society through the Sir Norman Lockyer Fellowship and by St Anne's
College, Oxford. The use of the Glamdring cluster of Oxford
University is acknowledged. I acknowledge the use of the package
\texttt{cosmomc}, available from \texttt{cosmologist.info},  and
the use of the Legacy Archive for Microwave Background Data
Analysis (LAMBDA). Support for LAMBDA is provided by the NASA
Office of Space Science.


\end{document}